# Biocompatibility of Additively Manufactured Fe-AZ31 Biodegradable Composites for Craniofacial Implant Applications

**Authors:** Xue Dong[1], Samuel Medina[1], Sai Pratyush Akula[2], Abby Chopoorian Fuchsman[1], Matthew W. Liao[1], Sophia Salingaros[1], Atieh Moridi[2], Jason A. Spector[1,3*]

* This is the corresponding author.

<u>Authors' institutional affiliations at which the work was carried out:</u>

[1] Laboratory of Bioregenerative Medicine & Surgery, Department of Surgery, Division of Plastic Surgery, Weill Cornell Medical College, New York, NY 10065, USA

[2] Sibley School of Mechanical and Aerospace Engineering, Cornell University, Ithaca, NY 14853, USA

[3] Nancy E. and Peter C. Meinig School of Biomedical Engineering, Cornell University, Ithaca, NY 14853, USA

**Corresponding Author:**

Jason A. Spector, MD, FACS

Chief, Division of Plastic and Reconstructive Surgery

Professor Plastic Surgery and Otolaryngology, Weill Cornell Medical College

Adjunct Professor, Nancy E. and Peter C. Meinig School of Bioengineering, Cornell University

Director, Laboratory of Bioregenerative Medicine and Surgery

Weill Cornell Medical College

525 East 68th Street, Payson 709-A

New York, NY 10065

646-962-8471 | <u>jas2037@med.cornell.edu</u>

ORCID: 0000-0002-4202-7151

**Abstract (<300 words):**

**Purpose:** Metallic plating systems composed of titanium and its alloys remain the standard treatment for craniofacial bony fixation but may require secondary removal due to infection, implant migration, or discomfort. Absorbable polymeric alternatives reduce those risks but lack sufficient strength for load-bearing applications. Thus, biodegradable metallic implants may eliminate complications and secondary procedures while maintaining structural integrity. Our previous work demonstrated the fabrication of immiscible Fe-AZ31 composites via additive manufacturing with improved degradation kinetics over pure Iron. This study aimed to evaluate the *in vitro* and *in vivo* biocompatibility of Fe-AZ31 composites for potential craniofacial fixation applications.

**Methods:** Pure iron (Fe), magnesium (Al-Mg-Zn) alloy (AZ31) and Fe-AZ31 samples were fabricated for extract-based cytotoxicity testing using HFF-1 fibroblasts, L929 fibroblasts and hFOB osteoblasts. Metal extracts were prepared at a 3 cm²/mL surface-to-volume ratio in complete media at 37°C and cell viability was measured by live/dead assay after 24 and 72 h exposure. For *in vivo* evaluation, Fe-AZ31, Fe, and titanium (Ti) plates were implanted subcutaneously in wild type mice for 6 weeks, 3 and 6 months. Implant degradation, histologic response, hematology, and serum biochemistry were assessed.

**Results:** Fe-AZ31 extracts demonstrated ≥ 70% cell viability across all cell types at both timepoints with normal cell morphology and adhesion, whereas AZ31 extracts caused marked cytotoxicity associated with pronounced alkalization (pH 10.53). *In vivo*, Fe-AZ31 implants exhibited gradual surface corrosion accompanied by mild, transient inflammation and minimal capsule formation over time. No systemic toxicity was observed. Hematology and serum biochemistry remained within physiological limits.

**Conclusion:** Additively manufactured Fe-AZ31 composites demonstrate acceptable cyto/biocompatibility and favorable tissue responses, supporting their development as bioresorbable metallic fixation devices for craniofacial reconstruction.

**Keywords (4-6):** Biodegradable metal; iron-magnesium composite; craniofacial reconstruction; cytotoxicity; *in vivo* biocompatibility

# 1. Introduction

Implantable metallic devices are utilized across surgical disciplines to provide structural stability and reliable fixation for fractured or reconstructed bone segments[1]. In craniofacial surgery, these devices are critical for achieving precise anatomical alignment, rigid fixation which promotes osseous union, and restoring both function and aesthetics after trauma, tumor resection or treatment of congenital deformity[2,3]. Most clinically utilized devices are fabricated from titanium and its alloys due to that metal's high specific strength, ductility, and corrosion resistance[4,5]. However, the long-term retention of non-absorbable metallic plates provides no functional advantage once bone healing is complete, typically 3 to 4 months after osteosynthesis, and can lead to complications such as infection, plate migration, palpability or chronic pain[6-9]. Especially in pediatric patients, the retained metallic hardware may further interfere with normal skeletal development and restrict craniofacial growth along suture lines, potentially resulting in secondary deformities[10]. Additionally, the stress shielding provided by the hardware may lead to pathologic bone resorption[11,12]. Removal of symptomatic hardware requires a revisional operation, which exposes patients to additional anesthesia, increased healthcare costs, and potential operative complications related to an additional surgery[13,14].

To address these limitations, bioabsorbable fixation devices have been developed to provide temporary mechanical support followed by gradual degradation, such as polymer-based polylactic acid (PLA) and polyglycolic acid (PGA), to eliminate the need for metal hardware removal[15-17]. These devices have been applied mostly in the pediatric setting for non-load bearing areas such as calvarial reconstruction but are restricted from broader application due to inadequate mechanical strength, unpredictable degradation kinetics and local inflammation caused by acidic degradation by-products[18,19]. In contrast, biodegradable metallic devices can address many of the shortcomings by combining high mechanical strength with controlled corrosion behavior. Iron (Fe), magnesium (Mg), and zinc (Zn)-based alloys are particularly promising because they degrade gradually under physiologic conditions and provide load-bearing strength comparable to-or in the case of Zn and Fe, exceeding-the typical range of cortical bone[20-22]. Although Fe exhibits excellent stiffness and ductility, it corrodes too slowly for clinical application, whereas Mg is too soft and degrades too rapidly, releasing hydrogen gas that can affect local tissue repair[23,24]. Zn demonstrates an intermediate corrosion rate between Fe and Mg and releases $Zn^{2+}$ ions during degradation, which can stimulate osteoblastic activity and support matrix mineralization, but its relatively limited ductility and poor fatigue resistance restrict its use in load-bearing applications[25,26].

To optimize degradation kinetics, Fe-Mg composites have been proposed that harness galvanic coupling between the Fe and Mg phases to accelerate Fe corrosion while stabilizing Mg dissolution[27,28]. The emergence of additive manufacturing (AM), particularly directed energy deposition (DED), has enabled the co-processing of otherwise immiscible alloys with controlled spatial arrangement of Fe and Mg-rich regions and refined microstructures[28,29]. Our previous work demonstrated that additively manufactured Fe-Mg alloy composites (Fe-AZ31; AZ31, is an alloy of magnesium with a typical composition of 3 wt% Al, 1 wt% Zn and 96 wt% Mg) achieved a retained Mg content of ~2.65 at% following DED processing. Comprehensive microstructural characterization using scanning electron microscopy (SEM) and energy-dispersive X-ray spectroscopy (EDS) confirmed uniform distribution of Mg-rich regions throughout the Fe matrix, while X-ray diffraction (XRD) verified the phases present and electron backscatter diffraction (EBSD) revealed grain refinement and phase distribution within the composite. Those features resulted in a corrosion rate approximately tenfold higher than pure Fe, while maintaining structural integrity and promoting formation of calcium- and phosphate-rich corrosion layers associated with favorable surface bioactivity. Those results suggest that AM-fabricated Fe-AZ31 may overcome the degradation-strength trade-off that limits current biodegradable metallic implants. Thus, in the current study we aimed to evaluate the biological safety and biocompatibility of AM-fabricated Fe-AZ31 composites via both *in vitro* and *in vivo* analyses to establish their potential for future craniofacial implant applications.

**2. Materials and Methods**

2.1. Design and Manufacturing of Metal Implants

As previously described, Fe-AZ31 composite (produced by mixing pure Fe and 10 at% of AZ31 Mg alloy powders in a low energy ball mill for 30 minutes) craniofacial fixation plates were custom manufactured using industrial grade powder-based directed energy deposition (DED) additive manufacturing (AM) (FormAlloy X2) under an inert atmosphere of <100 ppm of oxygen to prevent oxidation of samples during printing[30]. The dimensions, mass, and topography of manufactured Fe-AZ31 plates was designed to closely approximate those of the commercially available Stryker™ titanium craniofacial midface fixation plate (Ti). Fe-AZ31 plates (~0.6 mm in thickness, 10.0 mm in length and 2.0 mm in width) featured three equally spaced screw holes (~1.0 mm in diameter), mirroring the Stryker™ mini-plating design (# 55-08231). Plates composed of pure iron (Fe) were manufactured using the same techniques and design parameters. All the implants were low-temperature plasma sterilized with hydrogen peroxide before use.

2.2 *In Vitro* Cytotoxicity Assessment

Cytotoxicity of Fe-AZ31, Fe, and AZ31 samples was evaluated using an extract-based assay in accordance with ISO 10993-5 guidelines[31]. Plasma-sterilized plates were incubated in Dulbecco's Modified Eagle Medium (DMEM, 1X) supplemented with 10% fetal bovine serum and 1% penicillin-streptomycin) at a surface area-to-volume ratio of 3 cm²/mL at 37 °C for 24 h in capped tubes to obtain extracts. The extracts were collected, centrifuged to remove debris, and used immediately for testing. The pH of each extract was measured prior to cell exposure using a Thermo Scientific™ Orion Star™ A211 pH meter (Thermo Fisher Scientific, Waltham, MA, USA) with 1-2 mL of extract per material. Human dermal fibroblasts (HFF-1), human fetal osteoblasts (hFOB) and murine fibroblasts (L929) were obtained from the American Type Culture Collection (ATCC, Manassas, VA, USA), subcultured, and seeded at a density of 4,000 cells per well in 96-well plates (n=6 wells per group) and grown to 60-70% confluence prior to extract exposure. Cells were then exposed to material extracts for 24 and 72 h, with no medium exchange during the exposure period. Fresh medium, polyethylene (PE-positive control), and 70% ethanol (EtOH-negative control) were used, respectively. Cell viability was quantified using a Calcein AM/propidium iodide (PI) live/dead assay and imaged at 100× magnification using an Olympus 1X71 fluorescence microscope (Olympus, Tokyo, Japan).

2.3 Animal Model

All animal care and experimental procedures during this study followed the Guide for the Care and Use of Laboratory Animals and were approved by the Weill Cornell Medicine Institutional Animal Use and Care Committee (IACUC protocol # 2021-0015). Seventy-two C57BL/6 mice were randomly assigned to one of three experimental groups (Fe-AZ31, Fe, and Ti) with one implant placed per animal. Ti was used *in vivo* as the clinical standard control, whereas pure Mg-based AZ31 plates were not implanted because their rapid, gas-forming degradation is known to cause local hydrogen accumulation, mechanical instability, and soft-tissue disruption, which can confound long-term biocompatibility assessment and pose safety concerns in small animal models[32]. Implantation was performed under general anesthesia. The dorsal skin was shaved, prepped, and disinfected in a standard fashion. A small midline incision was made over the spine, and each plate was inserted into a subcutaneous pocket. Eight animals per group were euthanized after 6-week, 3-month, and 6-month post-implantation for analysis.

2.4 Determination of Change in Body Weight and Implant Mass

The mass of each plate was recorded prior to implantation and again at explantation. Any surrounding fibrous capsule developed was carefully removed before weighing to ensure accurate post-explantation measurements. Percent change in implant mass was calculated relative to the baseline pre-implantation values. Animal body weight was also measured prior to surgery and again at euthanasia, and percent change in body weight was calculated as above.

2.5 Blood Profiles

Peripheral blood samples were obtained via retro-orbital collection from each animal prior to implantation for baseline measurements, including complete blood count (CBC), serum biochemistry, and elemental Mg and Fe concentrations. Post-implantation blood samples were collected at the time of sacrifice at each time point (6 weeks, 3 months and 6 months) to monitor systemic responses. Whole blood was analyzed for hematologic parameters such as red and white blood cell counts, hemoglobin, hematocrit, platelet count, and differential leukocyte distribution. Serum samples were evaluated for renal and hepatic function markers (Blood urea nitrogen (BUN), creatinine, alanine aminotransferase (ALT), and aspartate aminotransferase (AST)), electrolyte balance (Na, K, Cl, P), and total protein and albumin concentrations. All analyses were performed by the Laboratory of Comparative Pathology (LCP) at Memorial Sloan Kettering Cancer Center (NY, NY) using standardized veterinary diagnostic protocols.

2.6 Histological analysis

Histological analysis was performed on tissue capsule samples surrounding each implant following explanation, as well as liver, kidney and spleen. Samples were fixed with 10% neutral-buffered formalin, processed, and embedded in paraffin using a Tissue-Tek VIP 6 Vacuum Infiltration Processor. Sections (10μm thick) were cut and stained with Hematoxylin and Eosin (H&E) to assess capsule organization, cellular infiltration, and the presence of metal-associated pigment or particulate debris. Capsule thickness was quantified by measuring the fibrous layer at two randomly selected regions per section under 40× magnification (n=8 per group). Immunofluorescence staining for CD86 and CD206 was performed on parallel sections to identify and evaluate macrophage populations. All slides were imaged under brightfield microscopy at 40×, 100×, 200× magnification using an Olympus 1X71 microscope (Olympus, Tokyo, Japan).

2.7 Statistical Analysis

All quantitative data are presented as mean ± standard deviation (SD). *In vitro* cell viability data were analyzed using two-way analysis of variance (ANOVA) followed by Tukey's post hoc test for multiple comparisons. For all *in vivo* outcomes, differences among material groups and timepoints were assessed using one-way ANOVA with Tukey's post hoc test. Statistical analyses were performed using GraphPad Prism 10.0 (GraphPad Software, San Diego, CA, USA). $P < 0.05$ was considered statistically significant.

**3. Results**

3.1 *In Vitro* Cytotoxicity

AZ31, Fe and additively manufactured Fe-AZ31 were fabricated and prepped for analysis. Scanning electron microscopy (SEM) revealed distinct surface morphologies among the three materials (**Figure 1A**). Fe exhibited a smooth, dense surface with minimal porosity, AZ31 displayed a sample surface with visible pores and oxide inclusions, and Fe-AZ31 showed a mixed microstructure containing Fe-rich and Mg-rich regions. Corresponding gross surface images showed matte silver for AZ31, metallic gray surfaces for Fe, and mottled metallic tones for Fe-AZ31 (**Figure 1B**).

Extract media collected after 24 h incubation at 37 °C exhibited distinct differences in appearance and pH among material groups (**Figure 1C, 2C**). The Fe-AZ31 extract appeared to be light orange, Fe extract was dark orange-brown, and AZ31 extract remained light pink (the color of control medium) but with visible dark precipitates. Quantitative pH measurements showed a pronounced alkalization in AZ31 extracts (pH 10.53), whereas Fe-AZ31 extracts showed only a mild elevation in pH (8.10), comparable to pure Fe extracts (pH 8.06), polyethylene control (pH 7.92), and fresh medium (pH 7.75).

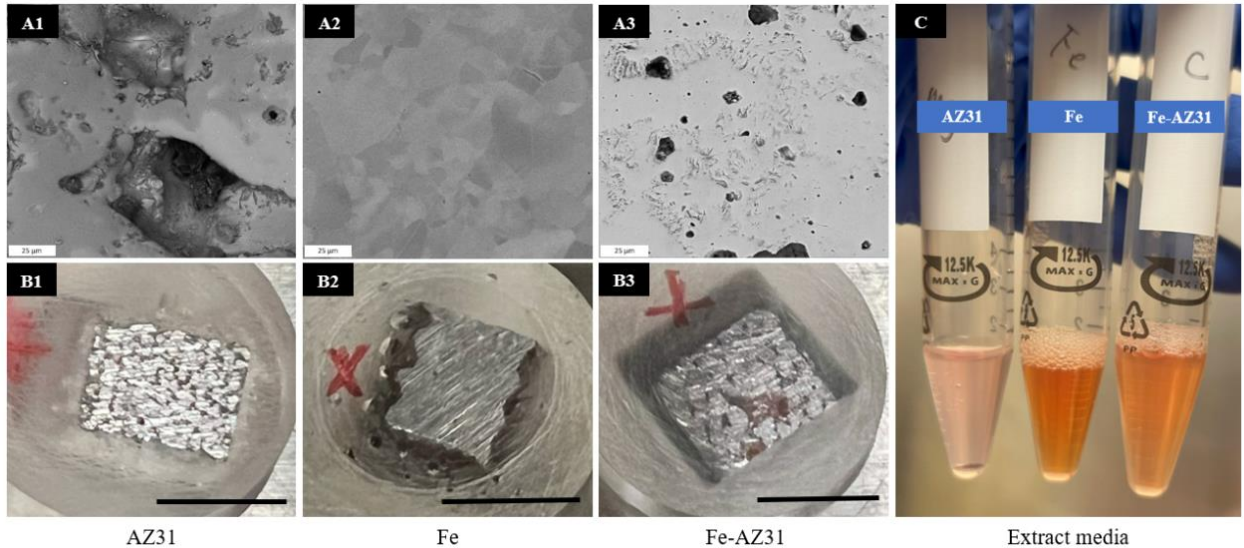

**Figure 1.** Fabrication and characterization of Fe-AZ31, Fe, and AZ31 plates for *in vitro* testing. A1-A3. Scanning electron micrographs showing distinct microstructures of AZ31, Fe and Fe-AZ31. AZ31 showed porosity along with oxide inclusions, Fe exhibited a smooth and dense surface, and Fe-AZ31 displayed a heterogeneous microstructure with Fe-rich and Mg-rich domains uniformly distributed. Scale bars = 25 μm. B1-B3. Gross images of plate surfaces showing matte silver AZ31, metallic gray Fe, and mottled metallic Fe-AZ31. Scale bars = 1.0 cm. C. Extract media prepared after 24 h incubation at 37 °C exhibited visible differences in color and debris formation. AZ31 extract remained light pink with dark precipitates, Fe extract was dark orange-brown, and Fe-AZ31 extract appeared to be orange.

At 24 h exposure, extract-based cytotoxicity testing demonstrated that Fe-AZ31 maintained high cell viability across all three cell lines (HFF-1: 89.5 ± 9.4%, hFOB: 73.9 ± 6.7%, L929: 81.7 ± 17.1%) (**Figure 2A, C**). Pure Fe extracts showed moderate reductions in cell viability (HFF-1: 82.4 ± 7.8%, hFOB: 75.6 ± 4.6%, L929: 74.8 ± 13.1%), while AZ31 extracts caused marked decreases, especially in HFF-1 and hFOB cells (HFF-1: 26.4 ± 10.0%, hFOB: 11.9 ± 5.5%, L929: 15.4 ± 4.0%), falling below the ISO 10993-5 threshold of 70% cell viability and indicating a cytotoxic response. Following 72 h exposure, Fe-AZ31 extracts continued to support cell viability above the ISO threshold (HFF-1: 88.6 ± 5.6%, hFOB: 72.5 ± 5.1%, L929: 77.5 ± 8.3%), whereas pure Fe extracts showed comparable but slightly reduced viability (HFF-1: 86.4 ± 7.6%, hFOB: 76.3 ± 3.7%, L929: 74.3 ± 11.9%). In contrast, AZ31 extracts exhibited near complete loss of viable cells at 72 h across all cell types (≤ 5% viability), confirming severe and progressive cytotoxicity. Statistically, AZ31 viability was significantly lower than all other groups across three cell lines (P < 0.05), whereas Fe and Fe-AZ31 showed no significant differences from each other (P > 0.05), but both showed moderately reduced viability when compared to the fresh medium and PE controls (P < 0.05). Morphologically, cells exposed to pure Fe and Fe-AZ31 extracts maintained normal spreading and attachment at both 24 and 72 h, whereas AZ31 extracts induced pronounced cell rounding, detachment, and loss of confluence (**Figure 2B**).

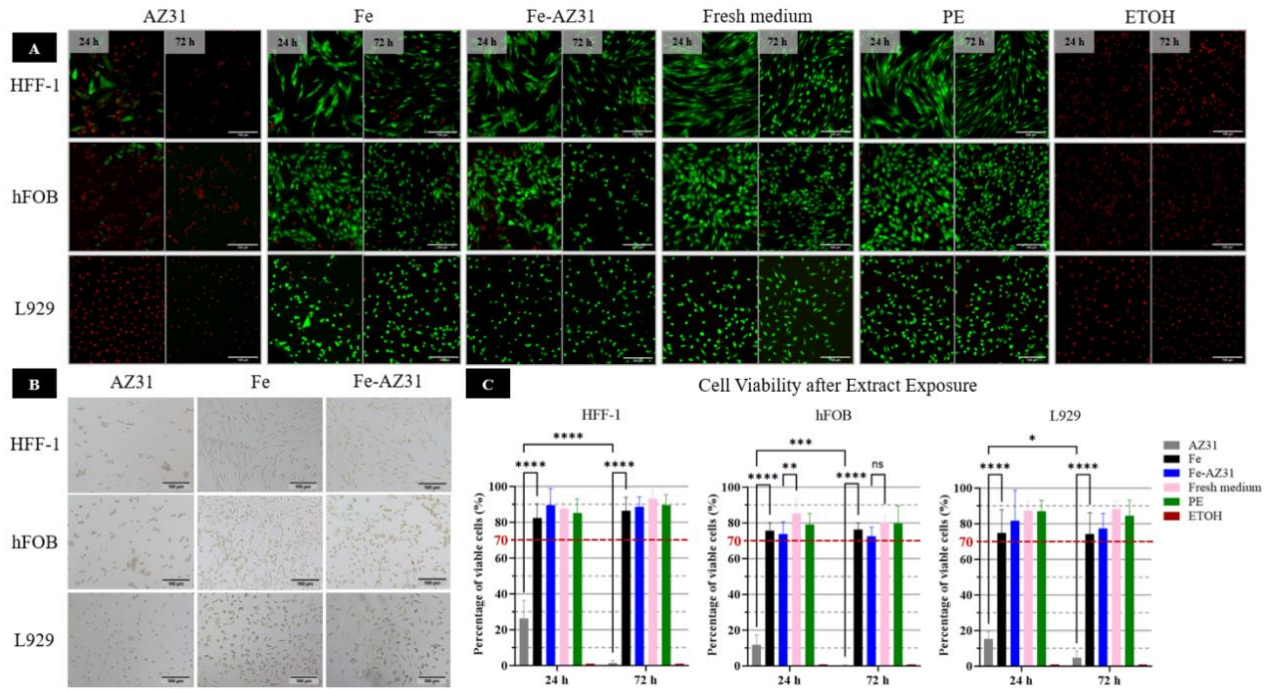

**Figure 2.** *In vitro* cytocompatibility of Fe-AZ31, Fe, and AZ31. A. Representative fluorescence images showing live (green, Calcein AM) and dead (red, propidium iodide (PI)) cells cultured in different extracts (HFF-1 fibroblasts, hFOB osteoblasts, and L929 fibroblasts after 24 and 72 h exposure to material extracts and controls). Scale bars = 100 μm. B. Representative brightfield images of cell morphology after 72 h exposure to material extracts. Cells in Fe-AZ31 and Fe extracts retained normal morphology and adhesion, whereas AZ31 extracts induced cell detachment and rounding. Scale bars = 100 μm. C. Quantitative cell viability analysis based on live/dead assays (n = 6). AZ31 extracts resulted in cell viability below the ISO 10993-5 threshold of 70% (dashed red line), indicating a cytotoxic response, whereas Fe-AZ31 and Fe extracts maintained viability above the 70% threshold after both 24 h and 72 h exposure. Statistical significance is indicated (*$P < 0.05$, **$P < 0.01$, ***$P < 0.001$, ****$P < 0.0001$; ns, not significant).

3.2 Body Weight and Implant Mass Changes

During the 6-month implantation, animal body weight increased gradually, with no significant differences observed at any timepoint ($P > 0.05$) (**Figure 3A**). Implant mass measurements showed material-dependent changes (**Figure 3B**). At 6 weeks, plates from all groups exhibited slight increases in measured mass compared to baseline, likely due to residual tissue that could not be fully removed at explantation; however, none of these variations in mass were statistically significant across timepoints ($P > 0.05$).

3.3 Histological Evaluation of Local and Systemic Responses

All animals recovered uneventfully after surgery, with no evidence of delayed wound healing or implant associated complications. At explantation, a thin fibrous capsule was observed surrounding all the implants (**Figure 3C**). The capsule formation appeared to be the most prominent and darkly pigmented around Fe plates, moderately brown around Fe-AZ31 plates, and thinnest and nearly transparent around Ti plates. The surrounding soft tissues appeared to be grossly normal without signs of inflammation, edema or necrosis.

Progressive surface changes were noticed in the Fe-based implants over time (**Figure 3C**). At 6 weeks and 3 months, both Fe and Fe-AZ31 plates exhibited mild surface darkening on both sides of each plate. By 6 months, Fe-AZ31 plates showed more extensive discoloration, and in a subset of samples, one surface appeared darker than the other and this variation was not related to implant orientation. Fe plates appeared uniformly brown and matte. In contrast, Ti plates retained their original metallic appearance over 6 months, with no visible changes.

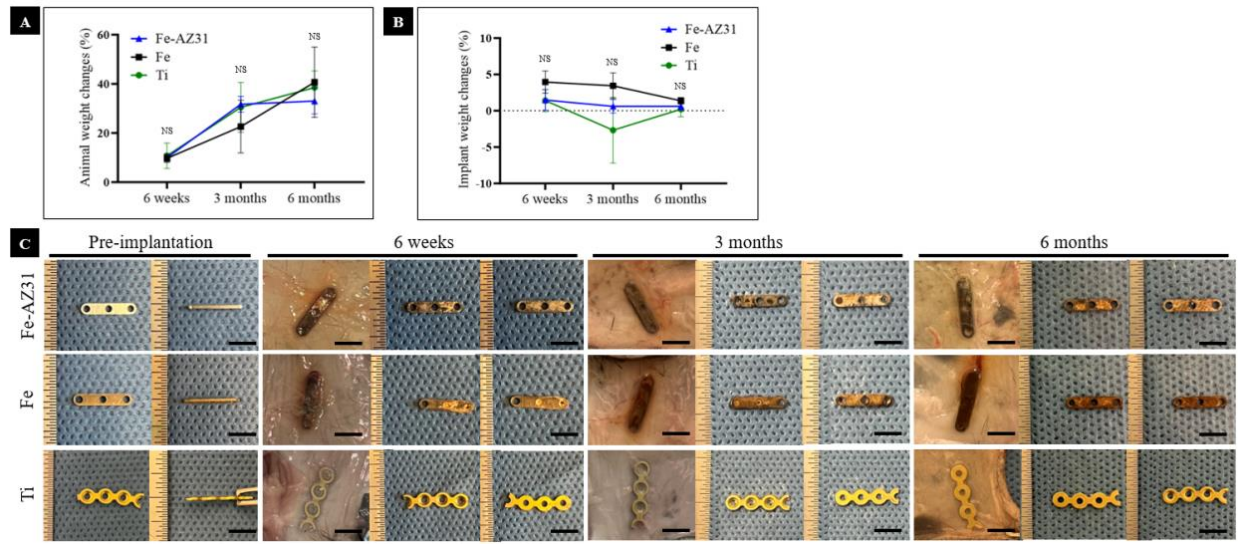

**Figure 3**. *In vivo* implantation of Fe-AZ31, Fe, and Ti plates. A. Body weight of animals over 6 months. B. Implant mass measurement over 6 months. C. Gross images of plates and surrounding capsule formation at pre-implantation and each explantation. Mild fibrous capsule formation was observed surrounding all implants, thicker around Fe and Fe-AZ31, and minimal around Ti. NS indicates no significant difference between the groups, p>0.05. Scale bars = 1.0 cm.

H&E staining of tissue capsules revealed thick fibrous layers surrounding both Fe-AZ31 and Fe implants at all timepoints (**Figure 4A**). Numerous dark pigment deposits of variable size were observed within these capsules, accompanied by increased cellular infiltration compared to Ti plates. Pigmented macrophages containing fine brown granules were occasionally observed adjacent to Fe-based implants at 3 and 6 months (**Figure 4B**). Immunofluorescence analysis demonstrated predominant CD206-positive M2 macrophages, whereas CD86-positive (M1) macrophages were minimal or undetectable at 6 months, and this finding was supported by quantitative IF analysis (**Figure 4D, E**). Notably, Fe implants exhibited significantly higher CD206-positive staining compared with Fe-AZ31 and Ti (P < 0.05). Capsule thickness measurements demonstrated that Fe implants consistently formed the thickest capsules of 689.8 ± 110.5μm at 6 weeks, 536.7 ± 198.0μm at 3 months and 529.8 ± 138.3μm at 6 months. Fe-AZ31 capsules were moderately thinner with 552.1 ± 153.1μm, 358.6 ± 114.5μm and 362.7 ± 135.5μm at the same timepoints. In contrast, Ti plates developed the thinnest fibrous layers with 433.2 ± 108.4μm at 6 weeks, 324.4 ± 144.6μm at 3 months and 289.1 ± 95.2μm at 6 months, which were significantly thinner than Fe plates across all timepoints (P < 0.05) (**Figure 4C**). In addition, histologic sections of the liver, kidney, and spleen from all the groups and time points were unremarkable, with normal architecture and no evidence of necrosis or pigment deposition (**Figure 4F**).

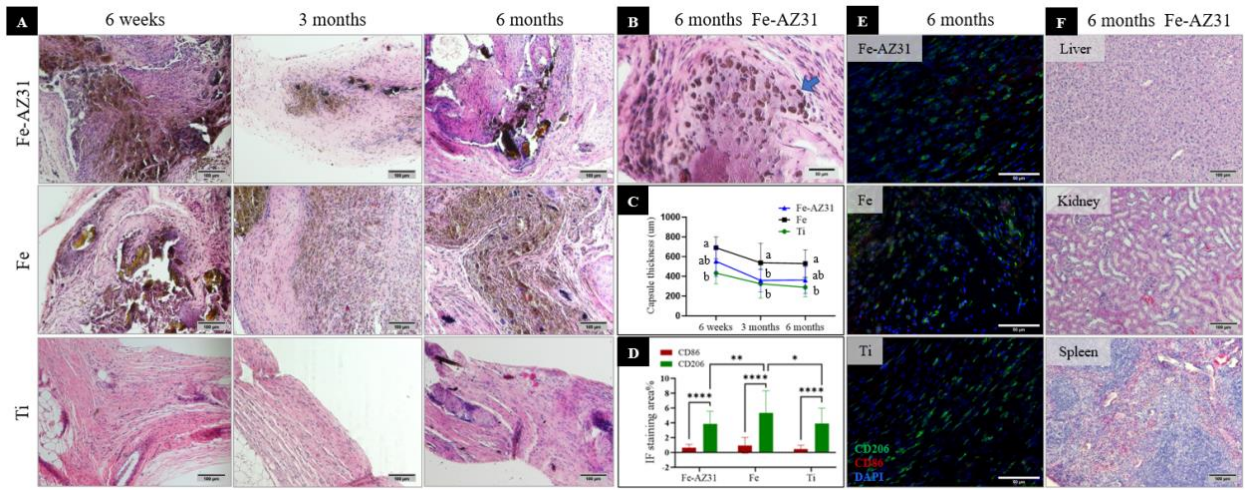

**Figure 4**. Histological analysis of implant capsules and distant organs. A. Representative H&E images of implant capsules at 6 weeks, 3 months, and 6 months. Increased cellularity and pigment deposition were evident in Fe-based groups. Scale bars = 100 μm. B. High-magnification H&E images showing metal-laden macrophages containing fine brown granules (arrow) near Fe-AZ31 implants at 6 months. Scale bars = 50 μm. C. Capsule thickness measurement of explants over 6 months. At each timepoint, groups sharing the same letter are not significantly different (P < 0.05). D. Quantitative immunofluorescence analysis of macrophage markers within implant capsules at 6 months. CD206-positive (M2) macrophages predominated over CD86-positive (M1) macrophages in all groups, with significantly higher CD206 staining area% in Fe implants compared with Fe-AZ31 and Ti (*P < 0.05, **P < 0.01, ****P < 0.0001). E. Representative immunofluorescent staining of CD206 in green and CD86 in red within implant capsules at 6 months. Scale bars = 50 μm. F. Representative H&E images of liver, kidney, and spleen sections showing normal tissue architecture and no evidence of necrosis or pigment deposition in Fe-AZ31 group at 6 months. Scale bars = 100 μm.

| Blood cell parameters | Pre-implantation | 6 weeks (Mean ± SD) | | | | 6 months (Mean ± SD ) (P value across time points) | | | | Reference range |
|---|---|---|---|---|---|---|---|---|---|---|
| | All groups | Fe-AZ31 | Fe | Ti | P value | Fe-AZ31 | Fe | Ti | P value | |
| RBC (M/uL) | 9.1 ± 0.5 | 9.5 ± 0.9 | 9.7 ± 1.0 | 8.9 ± 0.9 | 0.426 | 9.4 ± 0.8 (0.701) | 9.8 ± 0.5 (0.357) | 8.3 ± 2.4 (0.762) | 0.406 | 7.14 - 12.20 |
| HGB (g/dL) | 13.6 ± 0.6 | 13.9 ± 1.3 | 14.4 ± 1.4 | 13.1 ± 1.2 | 0.410 | 13.3 ± 1.2 (0.750) | 14.0 ± 0.8 (0.531) | 12.3 ± 3.5 (0.690) | 0.563 | 10.80 – 19.20 |
| HCT (%) | 47.3 ± 2.7 | 49.6 ± 3.9 | 51.5 ± 6.9 | 46.0 ± 3.7 | 0.337 | 46.1 ± 4.1 (0.423) | 48.6 ± 3.1 (0.449) | 42.8 ± 13.2 (0.724) | 0.621 | 37.20 - 62.00 |
| MCV (fL) | 52.1 ± 0.6 | 52.0 ± 1.7 | 52.6 ± 1.7 | 51.8 ± 1.8 | 0.781 | 49.1 ± 0.6 (0.005) | 49.6 ± 0.7 (0.008) | 50.9 ± 2.6 (0.645) | 0.303 | 42.60 - 56.00 |
| MCH (pg) | 15.0 ± 0.2 | 14.6 ± 0.1 | 14.8 ± 0.2 | 14.9 ± 0.3 | 0.274 | 14.2 ± 0.1 (0.0001) | 14.3 ± 0.2 (0.001) | 14.7 ± 0.6 (0.665) | 0.166 | 11.70 - 16.80 |
| MCHC (g/dL) | 28.8 ± 0.4 | 28.0 ± 1.0 | 28.1 ± 1.3 | 28.8 ± 0.5 | 0.453 | 28.9 ± 0.2 (0.148) | 28.7 ± 0.2 (0.377) | 29.0 ± 1.4 (0.955) | 0.904 | 24.60 - 35.90 |
| PLT (K/uL) | 476.0 ± 212.9 | 767.0 ± 162.3 | 661.3 ± 341.8 | 491.5 ± 237.0 | 0.355 | 846.5 ± 194.9 (0.05) | 852.3 ± 122.4 (0.146) | 608.3 ± 383.8 (0.783) | 0.356 | 565.00 - 2159.00 |
| WBC (K/uL) | 4.7 ± 2.2 | 7.7 ± 1.2 | 6.4 ± 2.2 | 6.9 ± 2.7 | 0.712 | 5.4 ± 1.7 (0.09) | 6.0 ± 1.4 (0.462) | 5.7 ± 1.8 (0.427) | 0.887 | 3.90 - 13.96 |
| Agranulocyte (%) | 89.6 ± 1.9 | 90.4 ± 6.3 | 90.2 ± 3.6 | 88.9 ± 2.1 | 0.872 | 85.7 ± 4.0 (0.332) | 87.3 ± 4.5 (0.481) | 84.2 ± 4.9 (0.09) | 0.651 | --- |
| LYMPH (%) | 83.7 ± 7.6 | 88.8 ± 6.7 | 87.8 ± 3.9 | 87.8 ± 1.8 | 0.941 | 84.7 ± 3.8 (0.506) | 86.3 ± 4.1 (0.587) | 83.0 ± 5.1 (0.426) | 0.578 | 61.26 - 87.12 |
| MONO (%) | 5.9 ± 6.4 | 1.6 ± 1.2 | 2.5 ± 3.0 | 1.1 ± 0.4 | 0.594 | 1.0 ± 0.2 (0.188) | 1.0 ± 0.6 (0.265) | 1.3 ± 0.3 (0.170) | 0.464 | 2.18 - 11.02 |
| Granulocyte (%) | 10.5 ± 1.9 | 12.2 ± 2.7 | 9.8 ± 3.6 | 11.1 ± 2.1 | 0.528 | 14.3 ± 4.0 (0.253) | 12.8 ± 4.5 (0.488) | 15.8 ± 4.9 (0.100) | 0.651 | --- |
| NEUT (%) | 9.1 ± 2.0 | 9.9 ± 3.2 | 7.9 ± 3.3 | 9.1 ± 2.1 | 0.629 | 12.3 ± 3.8 (0.356) | 11.3 ± 4.1 (0.380) | 14.4 ± 5.2 (0.09) | 0.606 | 7.36 - 28.59 |
| EO (%) | 1.1 ± 0.3 | 2.1 ± 0.6 | 1.7 ± 0.7 | 1.8 ± 0.6 | 0.703 | 1.8 ± 0.5 (0.056) | 1.3 ± 0.5 (0.317) | 1.2 ± 0.3 (0.075) | 0.223 | 0.13 - 4.51 |
| BASO (%) | 0.2 ± 0.3 | 0.1 ± 0.2 | 0.2 ± 0.1 | 0.3 ± 0.2 | 0.252 | 0.2 ± 0.2 (0.815) | 0.2 ± 0.1 (0.916) | 0.2 ± 0.2 (0.690) | 0.902 | 0.01 - 1.26 |

**Figure 5.** Complete blood counts (CBC) at pre-implantation, 6 weeks and 6 months. Red and white blood cell counts, hemoglobin, hematocrit, platelet levels, and differential leukocyte distributions remained within physiological ranges for all groups and time points (n = 4). Data are presented as mean ± SD; p < 0.05 was considered significant.

3.4 Hematology and serum biochemistry

Peripheral blood profiles remained within normal physiological ranges for all groups at each timepoint (**Figure 5**). No significant differences were observed in red or white blood cell counts, hemoglobin, hematocrit, or platelet levels compared to baseline. Most serum biochemistry parameters were comparable among groups or within reference range (**Figure 6A**). ALT and AST levels in Fe-based groups showed mild transient increases at 3 months compared to the Ti group, especially in Fe group, but all decreased and became comparable by 6 months. Electrolyte levels (Na and Cl), total protein and albumin concentrations showed no consistent group-specific variations. Potassium (K) levels decreased slightly in all the groups over time, with a significant difference at 6 months where Ti and Fe values were lower than Fe-AZ31 group. Phosphorus (P) levels were consistently lower in Fe-AZ31 group across all time points compared to Fe and Ti groups. Serum Mg and Fe concentrations showed minor fluctuations but no significant differences between groups, remaining stable throughout 6 months post-implantation (**Figure 6B, C**).

**4. Discussion**

The clinical need for absorbable fixation systems has increasingly driven interest in biodegradable metals as alternatives to permanent titanium hardware and polymer-based resorbable implants. While polymeric systems reduce the need for hardware removal, their mechanical limitations and inflammatory degradation byproducts constrain their use in load-bearing craniofacial applications[33,34]. Biodegradable metals can provide both mechanical strength and controlled resorption, but matching degradation kinetics to the bone-healing timeline remains difficult. This has led to growing interest in metal systems tailored to degrade in a controlled fashion while maintaining adequate early fixation.

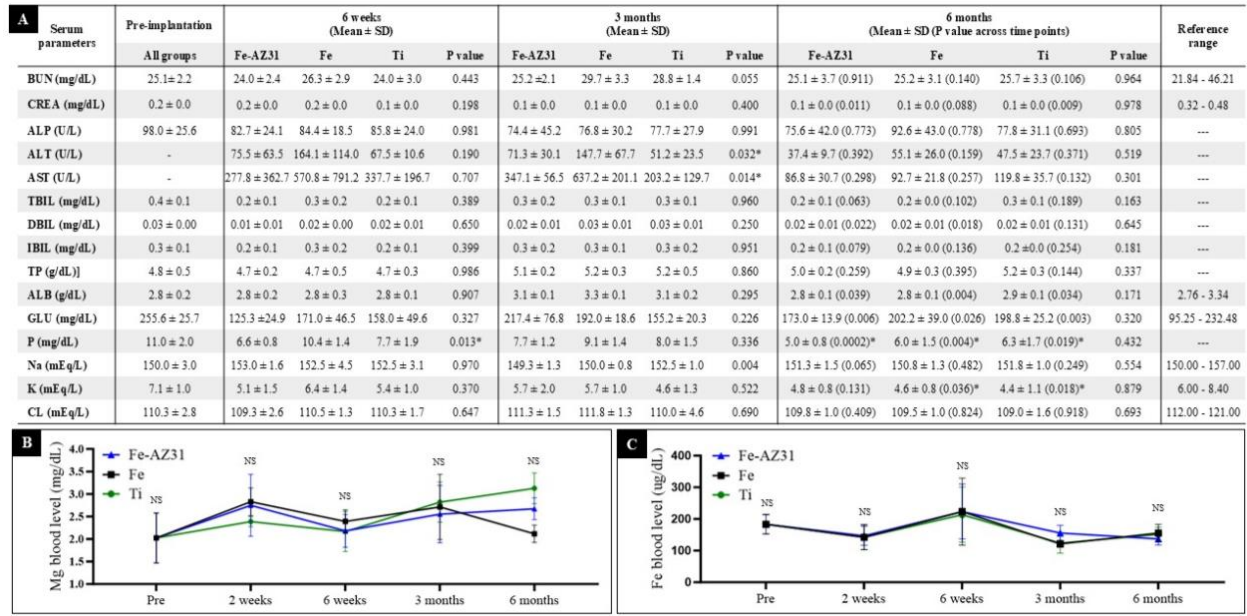

**Figure 6.** Serum biochemistry parameters at pre-implantation, 6 weeks, 3 and 6 months. A. Serum biochemistry analysis showed no long-term abnormalities across all groups (n = 4 per group per time point. B. Serum Mg concentration assessment. C. Serum Fe concentration assessment. Data are presented as mean ± SD; p < 0.05 was considered significant; NS indicates no significant difference between the groups, p>0.05.

Among degradable metals, Fe, Mg and Zn have received the most attention because they corrode under physiological conditions and provide mechanical properties closer to cortical bone than polymers[35]. Early clinical attempts to apply absorbable metals, mostly magnesium-based plates and screws, demonstrated some promising results in procedures such as mandibular fracture repair, hallux valgus correction and small bone stabilization[36,37]. Although theses implants supported early bone healing, their broader adoption was limited due to rapid and unpredictable corrosion, local alkalization and hydrogen gas pocket formation, often compromising mechanical stability or causing local tissue irritation. These shortcomings highlight the need for degradable metallic systems with more predictable and mechanically reliable performance. However, each of the three major degradable metal systems also has individual limitations. Fe corrodes at a very slow rate (~0.1 mm/year), Mg degrades too rapidly (~8 mm/year) accompanied by hydrogen gas evolution, and Zn, despite its intermediate corrosion behavior, lacks sufficient mechanical strength required for fixation[32,38-40]. To address these challenges, Fe-Mg bimetallic systems have recently been explored to couple the mechanical strength of Fe with the enhanced corrosion potential of Mg, thereby enabling more balanced degradation. Conventional manufacturing methods cannot produce refined Fe-Mg microstructures due to their intrinsic immiscibility, but additive manufacturing (AM), particularly directed energy deposition (DED), enables precise control of melt-pool dynamics and phase distribution[25,32,41]. Our previous work demonstrated that DED-fabricated Fe-AZ31 composites achieved Mg retention up to ~2.65 at% and exhibited a markedly accelerated corrosion rate compared to pure Fe, while maintaining structural integrity throughout degradation[30]. Comprehensive microstructural and compositional analyses using SEM, EDS, XRD, and EBSD confirmed a refined Fe matrix with uniformly distributed Mg-rich regions and grain size refinement. Importantly, immersion testing revealed a moderate pH increase for Fe-AZ31 compared with the steep alkalization observed for AZ31, along with the formation of calcium- and phosphate-containing corrosion products that may support bioactivity. The enhanced degradation kinetics of the composite were primarily attributed to galvanic interactions between dispersed Mg-rich sites and the surrounding Fe matrix, which act as localized corrosion initiation points and potentially suppress formation of a stable Fe passivation layer. Partial dissolution of Mg into the Fe matrix further alters the electrochemical potential, making the composite more electrochemically active and accelerating degradation. This study extends the evaluation by systematically assessing the biological performance of AM-fabricated Fe-AZ31 composites, focusing on cytocompatibility, *in vivo* tissue response, and systemic safety as critical steps toward translation into craniofacial fixation applications.

*In vitro* cytotoxicity testing demonstrated that Fe-AZ31 maintained high cell viability across human and mouse fibroblasts and osteoblasts, while AZ31 extracts caused significant cell detachment and death. This phenomenon has been well documented for Mg-based alloys and attributed to rapid Mg dissolution, which elevates local pH and disrupts extracellular ion balance, creating conditions that impair cell adhesion and survival[42-44]. In contrast, Fe-AZ31 extracts produced a chemical environment that preserved normal cell morphology and attachment, indicating that incorporation of Mg within the Fe matrix effectively mitigates the pH-driven cytotoxicity characteristic of pure Mg systems. Although moderate pH increases were observed across all the groups, including fresh medium and polyethylene controls, which is likely related to capped incubation conditions limiting $CO_2$ exchange, all extracts were prepared under identical conditions. Therefore, the pronounced alkalization observed in AZ31 extracts reflects intrinsic, material-dependent degradation behavior.

During *in vivo* implantation, Fe-AZ31 implants exhibited gradual surface corrosion without inducing tissue necrosis or excessive inflammation. Fe-based plates developed noticeably thicker fibrous capsules, while Fe-AZ31 plates showed a more moderate response, with thinner and more organized capsules compared to pure Fe. Both Fe groups exhibited mild macrophage infiltration and occasional pigment-containing macrophages, consistent with physiological phagocytic processing of iron oxide residues. Systemically, all major organs maintained normal histology without pigment deposition, and serum chemistry remained within physiological ranges. Mild transient elevations in ALT and AST at 3 months in Fe-based groups, especially Fe plates, may reflect short-term hepatic processing of corrosion byproducts, as no histologic abnormalities were observed in the liver. These fluctuation normalized by 6 months, likely reflecting hepatic adaptation to the iron load, and are consistent with previous reports describing temporary enzyme fluctuations during physiological clearance of Fe degradation products in animal models[45,46]. The consistently lower phosphorus levels in the Fe-AZ31 group compared with Fe and Ti may result from phosphate incorporation into surface corrosion products containing calcium and phosphate[32,47]. The slight reduction in potassium levels in Ti and Fe groups at 6 months likely reflects normal physiological variation rather than material-specific effects. Serum Fe and Mg concentrations remained stable and within physiologic parameters, indicating effective systemic regulation of released ions.

From a translational standpoint, Fe-AZ31 offers an advantageous balance between structural integrity and degradation behavior. The Fe ensures adequate stiffness during the early healing period, while the Mg phase accelerates corrosion toward the clinically relevant 3- to 6-month timeframe required for craniofacial osteosynthesis. Unlike pure Mg, Fe-AZ31 does not generate vast amounts of hydrogen gas and exhibits a more uniform corrosion pattern. Additive manufacturing further enhances its translational potential, enabling control over alloy composition, porosity, and architectural design to tailor degradation rate and mechanical retention to specific anatomical or patient needs.

As an initial step, this study was intentionally focused on biological feasibility, and mechanical characterization of the printed plates was therefore beyond its scope. Future studies will extend this work by incorporating quantitative evaluation of mechanical properties and corrosion kinetics to identify optimal compositions and processing parameters that achieve controlled (and faster) degradation rates (targeting approximately 1-3 mm per year *in vitro*), while maintaining sufficient strength to support bone healing for at least 4 months. In addition, extended *in vitro* investigations will evaluate longer-term cell-material interactions and osteogenic responses, including osteoblast function, mineralization, and lineage-specific marker expression, to better align biological assessment with craniofacial bone-healing processes. Moreover, because the subcutaneous implantation model used here does not recapitulate the mechanical and biochemical environment of bone healing, future investigations will therefore employ bone-specific, load-bearing models, such as mandibular or calvarial fixation, to evaluate degradation under functional stress, assess osseointegration, and explore the associated immune and tissue remodeling responses.

## 5. Conclusion

This study demonstrated that additively manufactured Fe-AZ31 bimetallic composites possess excellent *in vitro* cytocompatibility and favorable local and systemic tissue responses over 6 months *in vivo*. These findings establish the safety profile of Fe-AZ31 as a bioresorbable metallic material for temporary fixation, particularly in craniofacial reconstruction where controlled resorption and adequate mechanical support are required. Future optimization of composition, porosity, and processing parameters will evaluate and potentially enhance its mechanical performance and corrosion rate, facilitating translation toward load-bearing applications.

## 6. Declarations

Ethics Approval: All animal procedures were approved by the Institutional Animal Care and Use Committee of Weill Cornell Medicine (IACUC protocol #2021-0015) and conducted in compliance with the NIH Guide for the Care and Use of Laboratory Animals.

Funding: This research was funded by the Cornell Academic Integration Seed Grant. A.M. also acknowledges financial support from Johnson and Johnson Wistem2D scholar award.

Author Contributions: XD performed all experiments and wrote the manuscript. SM and MWL assisted with animal experiments and histology. AM conceived the project and supervised SPA in additive manufacturing. JAS conceived the project, provided resources, and oversaw all work. All authors reviewed and approved the final manuscript.

Competing Interests: The authors declare no competing financial or personal interests.

Data Availability: All data supporting this study are available from the corresponding author upon reasonable request.